\documentclass[review,12pt]{elsarticle}
\usepackage{graphicx}
\usepackage{graphics}
\usepackage{listings}
\usepackage{hyperref}
\usepackage{url}
\usepackage{amssymb}
\usepackage{siunitx}
\journal{Nuclear Inst. and Methods in Physics Research}

\begin{document}

\begin{frontmatter}

\title{Separation of electrons from pions in GEM TRD using deep learning}

\author[add1]{Nilay Kushawaha\fnref{fn1}}
\address[add1]{Department of  Physics, Indian Institute of Technology Indore, 
Madhya Pradesh-453552, India}
\ead{nilay.kushawaha112@gmail.com}

\address[add2]{Jefferson lab, 12000 Jefferson Avenue, Suite 15 Newport News, VA 23606}

\author[add2]{Yulia Furletova\fnref{fn1}}
\ead{yulia@jlab.org}

\author[add1]{Ankhi Roy\fnref{fn2}}
\ead{ankhi@iiti.ac.in}

\author[add2]{Dmitry Romanov\fnref{fn2}}
\ead{romanov@jlab.org}

\fntext[fn1]{First author}
\fntext[fn2]{Second author}
\begin{abstract}
Machine learning (ML) is not a completely new concept in the high energy physics community, in fact, many ML techniques have been employed since the early 80s to deal with a broad spectrum of physics problems. In this paper, we have presented a novel technique to separate electrons from pions in the Gas Electron Multiplier Transition Radiation Detector (GEM TRD) using deep learning. The Artificial Neural Network (ANN) model is trained on the Monte Carlo data simulated using the ATHENA based detector and simulation framework for Electron-Ion Collider (EIC) experiment. The ANN model does a good job of separating electrons from pions. 
\end{abstract}


\end{frontmatter}

\section{Introduction}
Particle identification is one of the major challenges in experimental physics. The identification of a stable particle is done either on the basis of their interaction or by determining their masses. In traditional particle physics experiments, particles are identified by the characteristic signature they leave in the detector. Conventionally, particle identification was done using the cut-based method where a threshold was fixed and if the signature of the particle was more than the threshold value, then it was classified as a signal. With the advancement of superior hardware and smart algorithms, various machine learning and deep learning techniques came into existence. Deep learning \cite{geron} and Artificial Neural Networks \cite{ANNbook} have become the most popular tool for research, data-driven and prediction based applications. \\
Transition Radiation Detectors (TRDs) are used for electron identification and for electron/hadron separation (in addition to calorimeter $\&$ ring-imaging Cherenkov detector) in some particle physics experiment \cite{yellowReport}.\\
In this paper, we will discuss the Gas Electron Multiplier Transition Radiation Detector(GEM TRD) \cite{yuliaGemtrd}, simulation of GEM TRD and the deep learning technique to separate electrons from pions. 
\section{Software Implementation of Detector Setup}
\subsection{Physics Processes}
Transition radiation (TR) \cite{yuliaGemtrd} is produced by charged particles when they cross the boundary between two media with different dielectric constants. When electrons travel through the radiator, TR photons are produced. The total TR energy is proportional to the $\gamma$-factor \cite{yellowReport} of the charged particle. Some TR photons are absorbed in the 3 cm gas volume (Xe-based mixture) of GEM. The X-ray TR photons are extremely forward peaked, and therefore their clusters overlap with the $\frac{dE}{dx}$ of charged particle \cite{leoBook}.\\
In the case of pions, no TR photons are produced. Pions begin to produce TR at energies greater than 100 GeV. For the EIC experiment, the particles will be produced in the energy range of up to 50 GeV, therefore for our simulation we used particles with energy, E $\sim$ 6 GeV \cite{yellowReport}.  
\subsection{Software Dependencies}
The ATHENA singularity container \cite{singularity} contains all the necessary softwares required for the construction, simulation, visualization of the detector as well as particle generation, analysis, and reconstruction.\\
DD4hep \cite{dd4hep} is a software framework included in the singularity container for providing a complete solution to full detector description (geometry, materials, visualization, readout, alignment, calibration,etc.) for the full experiment life cycle which includes detector concept development, detector optimization, construction, operation. \\
For simulating the data we are using the “ddsim” \cite{ddsim} simulation package provided by the ATHENA singularity container \cite{singularity}. We are simulating the data for electrons and pions in two separate root files. More information about the data is discussed in section \ref{data_gen}.
\subsection{Detector Simulation using DD4hep Software}
\begin{figure*}[!ht]
\minipage{0.32\textwidth}
  \includegraphics[width=\linewidth]{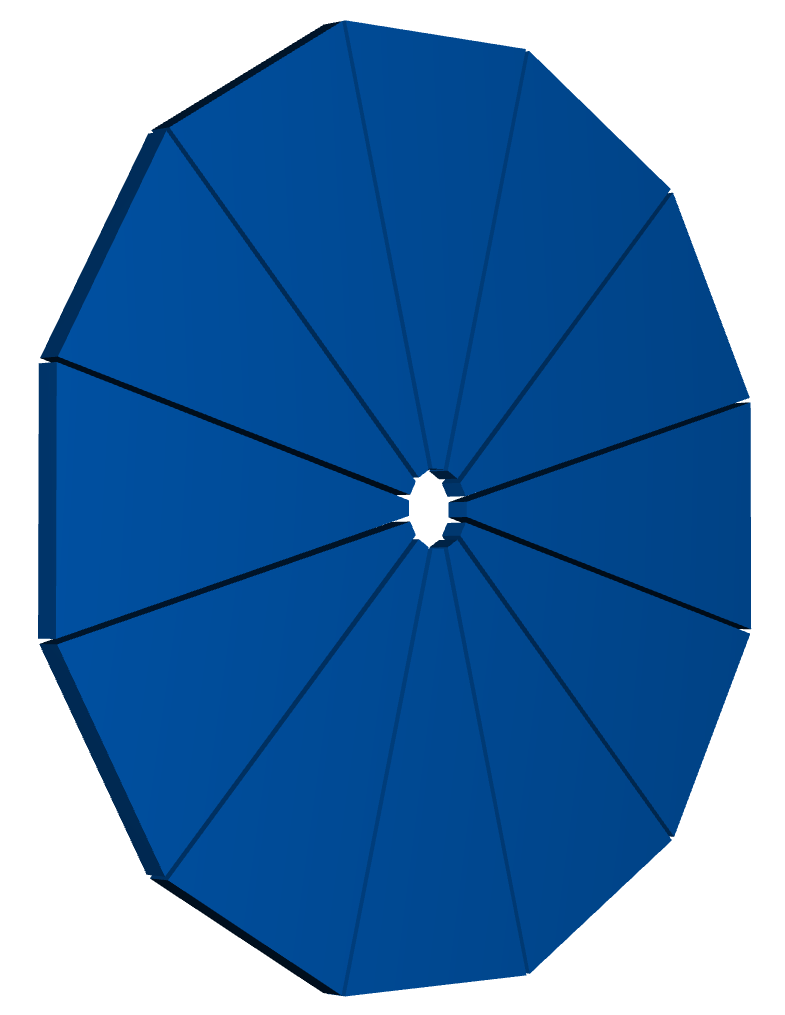}
\endminipage\hfill
\minipage{0.32\textwidth}
  \includegraphics[width=\linewidth]{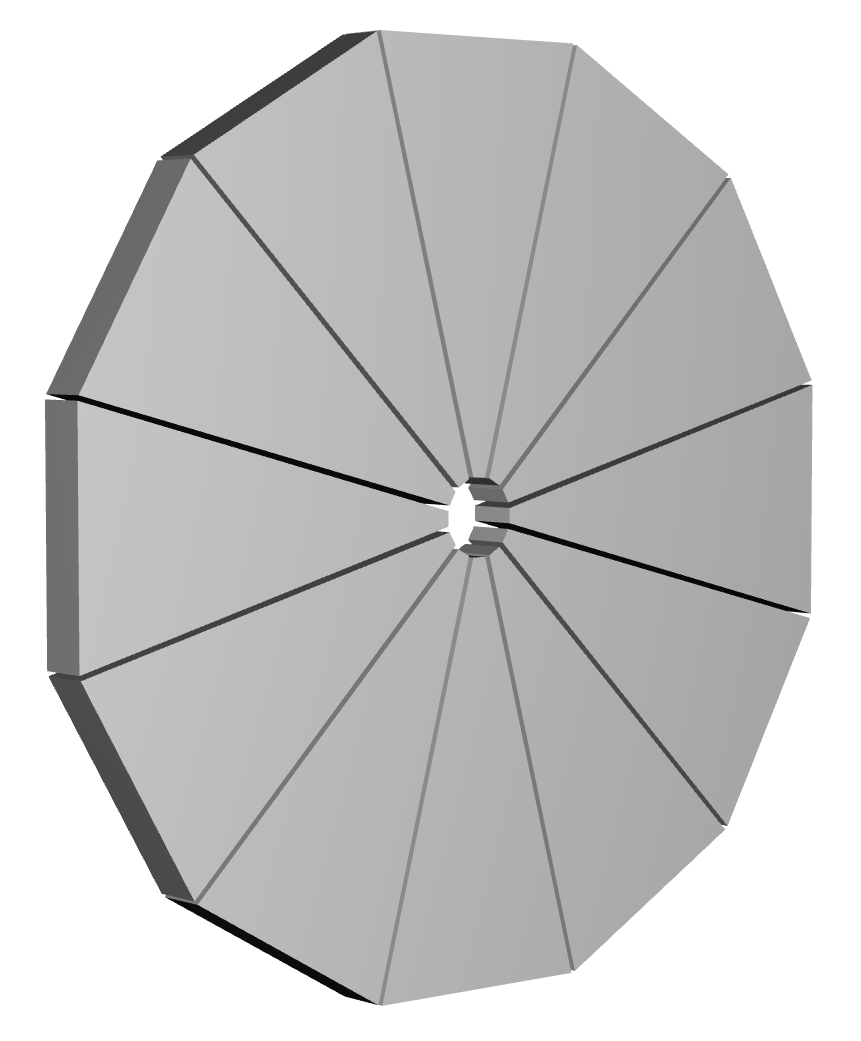}
\endminipage\hfill
\minipage{0.32\textwidth}%
  \includegraphics[width=\linewidth]{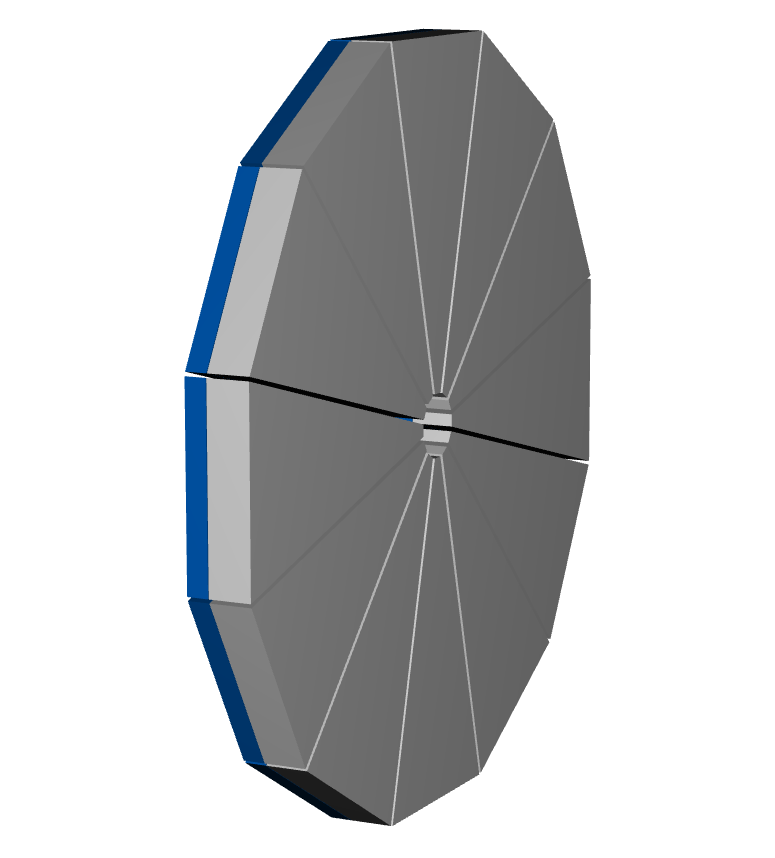}
\endminipage
\caption{\textit{left : GEM TRD module using DD4hep software with thickness 30 mm, middle : Radiator module using DD4hep software with thickness 150 mm, right : GEM TRD and radiator with a gap of 200 microns}}\label{fig:gemRad}
\end{figure*}

DD4hep software is used to create the detector and radiator geometries. Two disk-like shapes are created: one for the TR-radiator and one for GEM. The GEM disc has a thickness of 3 cm, and the material inside it is composed of xenon gas (Xe) and carbon dioxide (CO$_{2}$) in the 80:20 ratio, the thickness of the radiator is 15 cm and is enclosed with thin sheets of mylar foil (CH$_{2}$ \& Air). Figure \ref{fig:gemRad} right shows the radiator along with the GEM TRD. The radiator and the GEM are separated by a gap of $\sim$ \SI{200}{\micro\metre} filled with air. \\
To set up the physics list for the sensitive GEM layer and the TR-radiator, we use the QSGP$\_$BERT reference physics \cite{physicsList}, which includes all relevant physics processes for particles with energies below 10 GeV. 
\subsection{Data Generation $\&$ Overall Setup}
\label{data_gen}
We are simulating 1000k records each for electrons and pions in two separate root files using a 6 GeV particle gun for training the model. For testing the performance of the model we are simulating 500k mixed records. The root files provide information about the particle's position coordinate, energy deposit, drift time, and momentum information. In order to create features for the machine learning model, the Z-coordinate (Position Z) of the particles in the drift region is split into 69 bins with the corresponding energy deposit associated with it as the features.\\
\begin{figure}[htp]
    \centering
    \includegraphics[width=10cm,height=7cm]{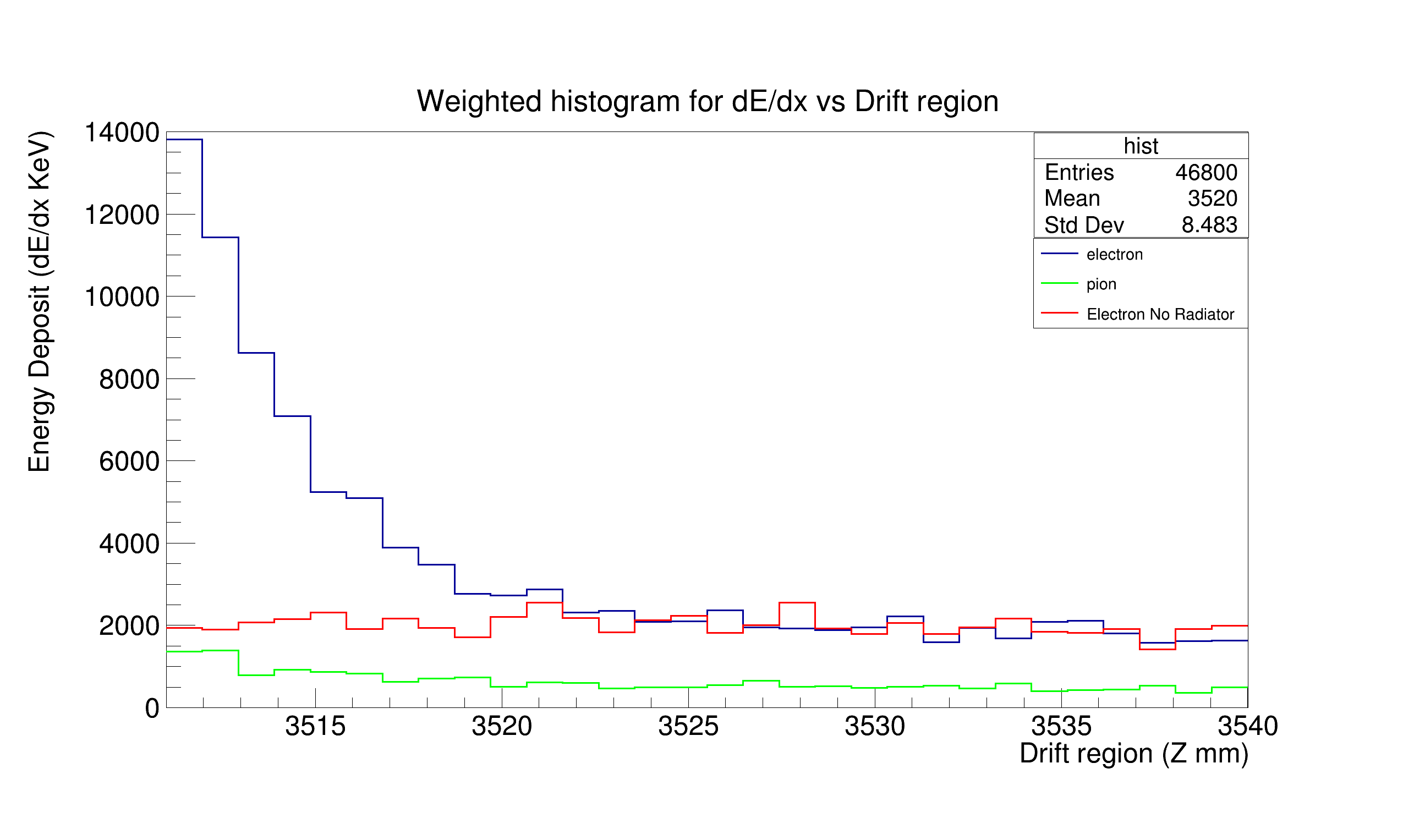}
    \caption{\textit{DD4hep simulation of $\frac{dE}{dx}$+ TR photons vs drift distance for electrons with and without radiator (blue, red) and pions (green)}}
    \label{fig:hist2d}
\end{figure}

We plot a 1-D weighted histogram of the Z-coordinate for both electrons and pions in the 30 mm drift gap with the energy deposit ($\frac{dE}{dx}$) as the weight parameter as shown in figure \ref{fig:hist2d}. Particles enter the GEM from the left at value 3510 mm and the number corresponding to 3540 mm is the readout of the detector setup. The energy deposit of the electrons is initially quite high whereas the pion's energy deposit remains flat throughout the drift distance. \\
When electron enters the radiator, it interacts with the material and generates soft and hard TR photons. The soft TR photons are mostly absorbed near the entrance window of the sensitive volume, resulting in the enhancement seen in the graph in case of electrons. Some hard TR photons are absorbed along the drift volume of GEM or could leave the volume undetected. However, in the absence of radiator the electrons also have a flat energy deposit throughout the drift distance as shown in figure \ref{fig:hist2d} in red color. Note, that the $\frac{dE}{dx}$ of electrons without the radiator is higher than for pions.
\section{Neural architecture and data generation}
\subsection{Working of Artificial Neural Network}
Artificial Neural Networks (ANN) \cite{ANNbook} have been around for quite a long time, they have been studied for many years in the hope of achieving human-like performance. ANN is proven to be a powerful tool to find and understand the uniqueness of certain features of the data which are invisible to human eyes, look for hidden patterns and features in the data \cite{ann_biology}.\\
The working of an ANN can simply be understood as the mapping of function from one space to some other space either linearly or using some higher-order relation. It can be used for solving both classification as well as regression problems. The ANN described in our paper consists of three layers:
\begin{itemize}
    \item Input layer: It takes the input vectors from the user and multiplies the respective branch weights to it.
    \item Hidden layer: The hidden layer is a collection of neurons that perform all computations on the input data. It is responsible for learning complex patterns from the data.
    \item Output layer: It gives the final predicted output value based on the input features.
\end{itemize}
The neural network model takes the input, multiplies some branch weight to it, concatenates them and applies an activation function on top of it. The prediction of neural network model is compared with the true label and the loss function is calculated. The loss function is then minimized using an optimizer \cite{geron} by tweaking the weights until we reach the global minima or a considerable loss value.
\subsection{Model Architecture}
\begin{figure}[htp]
    \centering
    \includegraphics[width=14.3cm,height=1.3cm]{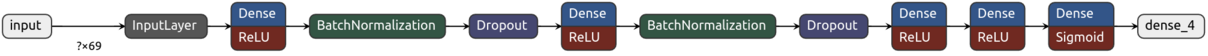}
    \caption{\textit{Artificial Neural Network architecture for the model with 69 features}}
    \label{fig:modelArchi}
\end{figure}

We have designed an ANN model with one input layer, four hidden layers and an output layer using the Keras framework \cite{keras}. Figure \ref{fig:modelArchi} shows the complete ANN architecture along with the various parameters associated with it. 
The input layer contains 69 nodes or neurons to handle the input vectors. The hidden layers contain 500, 300, 200 and 100 neurons sequentially with "RELU" activation function \cite{relu}. The presence of dropout layer \cite{geron} and batch normalization \cite{batch_norm} reduces overfitting of the model on the training data. In the output layer, we have one neuron with sigmoid activation function. The loss function we are using is binary cross entropy \cite{keras} as we have a binary problem statement. To decrease the loss function we are using the "ADAM" optimizer \cite{adam}. The ANN model is trained for 100 epochs and the predictions are made on the test data.
\subsection{ANN Performance on Test Data}
\begin{figure*}[!ht]
\minipage{0.32\textwidth}
  \includegraphics[width=\linewidth]{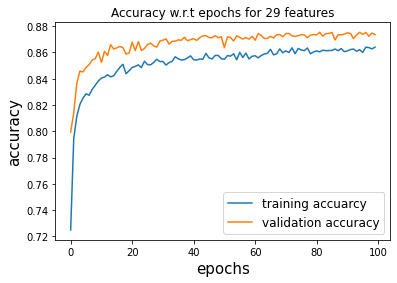}
\endminipage\hfill
\minipage{0.32\textwidth}
  \includegraphics[width=\linewidth]{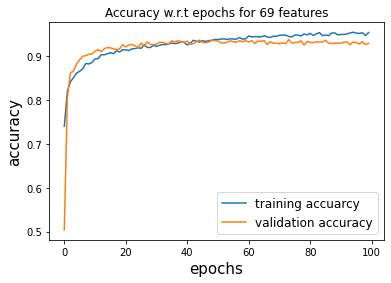}
\endminipage\hfill
\minipage{0.32\textwidth}%
  \includegraphics[width=\linewidth]{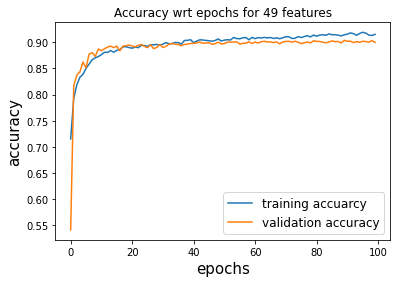}
\endminipage
\caption{\textit{Improvement in accuracy with respect to epochs for training and validation data, left: 29 features, middle: 69 features, right: 49 features}}\label{fig:accuracy_ann}
\end{figure*}

We trained three distinct ANN models to compare the performance of the model on the test data. The first model has 29 features, the second has 49, and the third model has 69. However, when the predictions of all three models were compared, it was discovered that the model with 69 features outperformed the other two.
\begin{figure*}[!ht]
\minipage{0.32\textwidth}
  \includegraphics[width=\linewidth]{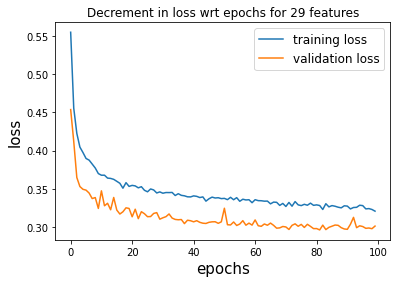}
\endminipage\hfill
\minipage{0.32\textwidth}
  \includegraphics[width=\linewidth]{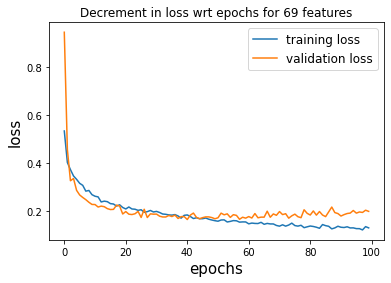}
\endminipage\hfill
\minipage{0.32\textwidth}%
  \includegraphics[width=\linewidth]{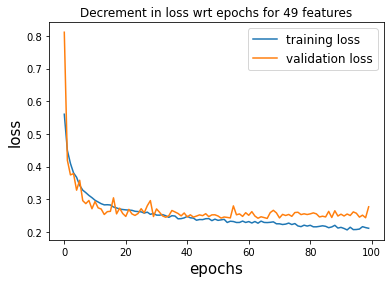}
\endminipage
\caption{\textit{Change in loss value with respect to epochs for training and validation data, left: 29 features, middle: 69 features, right: 49 features}}\label{fig:loss_ann}
\end{figure*}
The loss function decrement with respect to epochs for the three models is shown in figure \ref{fig:loss_ann}. We can see that the loss value in middle figure is close to zero, both for training and validation. The accuracy of all the three models with respect to epochs is shown in figure \ref{fig:accuracy_ann}. 
Again it can be seen that the model with the greater number of features has larger accuracy. Table \ref{tab:myTab} shows the accuracy and F1 score \footnote{F1 score is the harmonic mean of precision \cite{f1_score} and recall \cite{pr}. It is a special case of F-beta score \cite{geron}.} for the three ANN models.\\
\begin{table*}[]
    \centering
    \begin{tabular}{ |p{3.2cm}|p{1.5cm}|p{1.5cm}|p{1.5cm}|}
    \hline
    Model &Accuracy &F1 score (0) &F1 score (1) \\
    \hline
    ANNmodel29bins & 0.87     & 0.87         & 0.87 \\
    \hline
    ANNmodel49bins & 0.90     & 0.90         & 0.90 \\
    \hline
    ANNmodel69bins & 0.93     & 0.93         & 0.93\\
    \hline
    \end{tabular}
    \caption{\textit{Accuracy and f1 score for three different models where (1) refers to electrons and (0) refers to pions}}
    \label{tab:myTab}
\end{table*}
We can observe that the ANN model with 69 features beats all other models; however, we cannot raise the number of bins or features beyond 69 since the detector setup's electronics have a threshold resolution on the number of bins it can work with. The existing electronics can only handle 30 bins, but we have provided 69 bins in this study to illustrate that increasing the number of bins can lead to more efficient separation of electrons from pions, providing space for improvement within the detector setup's electronics. 
\section{Results and discussions}
\begin{figure*}[!ht]
\minipage{0.32\textwidth}
  \includegraphics[width=\linewidth]{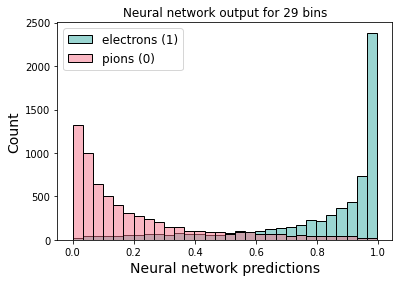}
\endminipage\hfill
\minipage{0.32\textwidth}
  \includegraphics[width=\linewidth]{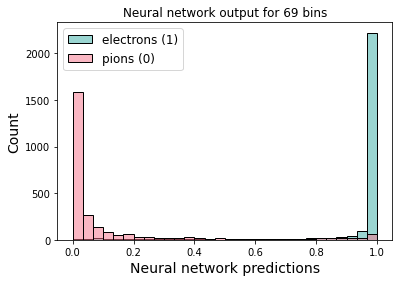}
\endminipage\hfill
\minipage{0.32\textwidth}%
  \includegraphics[width=\linewidth]{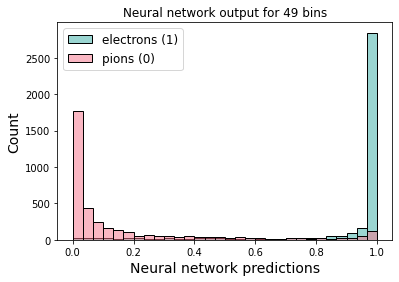}
\endminipage
\caption{\textit{Signal vs background categorization for electrons and pions, left: 29 features, middle: 69 features, right: 49 features}}\label{fig:signalBack}
\end{figure*}

\begin{figure*}[!ht]
\minipage{0.32\textwidth}
  \includegraphics[width=\linewidth]{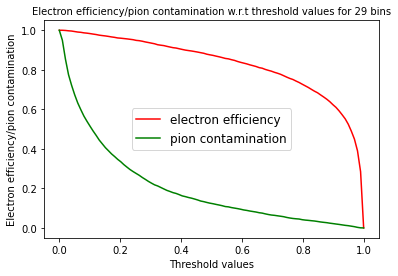}
\endminipage\hfill
\minipage{0.32\textwidth}
  \includegraphics[width=\linewidth]{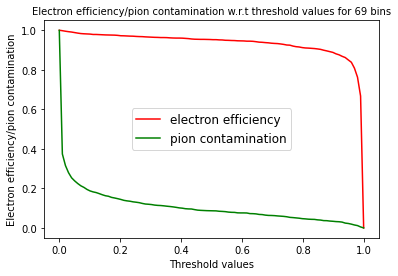}
\endminipage\hfill
\minipage{0.32\textwidth}%
  \includegraphics[width=\linewidth]{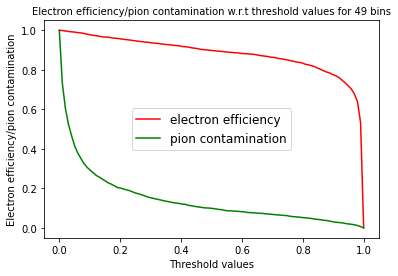}
\endminipage
\caption{\textit{Electron efficiency/pion contamination w.r.t threshold values for three different ANN models, left: 29 features, middle: 69 features, right: 49 features}}\label{fig:elepion}
\end{figure*}

Figure \ref{fig:signalBack} shows the output  of the Artificial Neural Network for all three models. 
We construct a table \ref{tab:thresholdtab} with the electron efficiency and pion rejection factor for the different threshold values in order to set an acceptable cut on the output of the ANN model. We can simply identify a balance between the two parameters (electron efficiency and pion rejection factor) according to our needs.\\
For example : if we put the threshold cut at around 0.6 for the ANN model with 69 features, we will get an electron efficiency of 93$\%$ and a pion rejection factor of 15.
The tradeoff between electron efficiency and pion contamination w.r.t the various threshold values is shown in figure \ref{fig:elepion}. 
\begin{table*}[]
    \centering
    \begin{tabular}{ |p{3.2cm}|p{2.3cm}|p{2.3cm}|p{2.3cm}|}
    \hline
    Model &Threshold value &Electron efficiency &Pion rejection factor \\
    \hline
                  & 0.2     & 0.958         & 3.171 \\
                  & 0.4     & 0.895         & 6.616 \\
    ANNmodel29bins & 0.6     & 0.829         & 11.670 \\
                  & 0.8     & 0.712         & 27.282 \\
                  & 0.9     & 0.594         & 59.663 \\
    \hline
                  & 0.2     & 0.952         & 5.466 \\
                  & 0.4     & 0.917         & 8.871 \\
    ANNmodel49bins & 0.6     & 0.880         & 12.6 \\
                  & 0.8     & 0.823         & 21.865 \\
                  & 0.9     & 0.758         & 33.486 \\
    \hline
                  & 0.2     & 0.976         & 7.266 \\
                  & 0.4     & 0.957         & 10.368 \\
    ANNmodel69bins & 0.6     & 0.938         & 15.613 \\
                  & 0.8     & 0.908         & 24.287 \\
                  & 0.9     & 0.891         & 30.149\\
    \hline
    \end{tabular}
    \caption{\textit{Electron efficiency and pion rejection factor for the three ANN models with different threshold values}}
    \label{tab:thresholdtab}
\end{table*}
\section{Conclusion}
Electron identification will be very important for the future Electron-Ion Collider (EIC) experiment due to the expected large hadron background. \\
The GEM TRD module with 30 mm drift gap along with the 150 mm radiator provides a desirable $e$/$\pi$ separation. The results presented from the DD4hep simulation of the GEM TRD setup and the radiator show that at 92$\%$ electron efficiency, a pion rejection factor of about 8.9 can be achieved for the ANN model with 49 bins. The electron efficiency and the pion rejection factor can further be increased by using an ANN model with greater number of bins/features.
\section{Acknowledgements}
This work was supported by the U.S. Department of Energy, Office of Science, Office of Nuclear Physics under contract DE-AC05-06OR23177.

\end{document}